\documentclass[twocolumn,showpacs,preprintnumbers,amsmath,amssymb]{revtex4}
\usepackage{graphicx}% Include figure files
\usepackage{dcolumn}% Align table columns on decimal point
\usepackage{bm}% bold math
\begin{document}

%\preprint{APS/123-QED}

\title{DNA nanotweezers studied with a coarse-grained model of DNA}
\author{Thomas E. Ouldridge$^1$}
\author{Ard A. Louis$^1$}
\author{Jonathan P. K. Doye$^2$}
\affiliation{$^1$Rudolf Peierls Centre for Theoretical Physics, 1 Keble
        Road, Oxford, UK OX1 3NP, UK \\ 
        $^2$Physical \& Theoretical Chemistry Laboratory, Department of Chemistry, University of Oxford, 
  South Parks Road, Oxford, OX1 3QZ, UK}

\date{\today}

\begin{abstract}
We introduce a coarse-grained rigid nucleotide model of DNA that reproduces the basic thermodynamics of short strands: duplex hybridization, single-stranded stacking and hairpin formation,  and also captures the essential structural properties of DNA: the helical pitch, persistence length and torsional stiffness of double-stranded molecules, as well as the comparative flexibility of unstacked single strands.  We apply the model to calculate the detailed free-energy landscape of one full cycle of 
 DNA `tweezers', a  simple  machine driven by hybridization and strand displacement. 
 \end{abstract}

%34.20.-b 	Interatomic and intermolecular potentials and forces, potential energy surfaces for collisions
%34.20.Gj 	Intermolecular and atom-molecule potentials and forces
%81.07.-b 	Nanoscale materials and structures: fabrication and characterization
%81.07.Nb 	Molecular nanostructures
%81.16.-c 	Methods of micro- and nanofabrication and processing 
%81.16.Dn 	Self-assembly
%82.39.-k 	Chemical kinetics in biological systems
%82.39.Pj 	Nucleic acids, DNA and RNA bases (for DNA, see 87.14.gk; for RNA, see 87.14.gn)
%87.14.-g 	Biomolecules: types
%87.14.gk 	DNA
%87.15.-v 	Biomolecules: structure and physical properties
%87.15.A- 	Theory, modeling, and computer simulation
%87.15.ak 	Monte Carlo simulations
%87.15.H- 	Dynamics of biomolecules
%87.15.hp 	Conformational changes
%87.15.N- 	Properties of solutions of macromolecules
% PACS, the Physics and Astronomy  Classification Scheme.
\maketitle

The field of DNA nanotechnology has grown rapidly in recent years as investigators have harnessed the  selectivity of DNA base pairing to form many different kinds of structures. Recent examples include: large ribbons \cite{Yan2003}, two dimensional lattices \cite{Malo2005} and polyhedra \cite{Seeman2003, Goodman2005}, made by hybridizing systems of short strands (oligonucleotides). Another technique, DNA origami \cite{Rothemund06}, uses short `staple' strands to fold a long polynucleotide into almost any two-dimensional shape, and has recently been extended to three-dimensional structures \cite{Andersen09,Douglas09}.

The free energy of hybridization can also be harnessed in artificial molecular machines. The simplest designs, such as DNA `tweezers' \cite{Yurke2000}, use alternating addition of two complementary strands to drive a system through a conformational cycle. 
More sophisticated autonomous machines catalyze the hybridization of strands initially present in inert forms, such as hairpins. Hybridization cycles can be coupled to a DNA track, creating DNA walkers that undergo directional motion \cite{Green2008,Omabegho2009}.

Computer simulations of these DNA nanosystems would provide highly desirable insight into the details of the processes involved in assembly or mechanical cycles. Unfortunately, the system sizes and time scales involved make all-atom simulations prohibitively expensive. Instead, models that coarse-grain out the microscopic details must be used.
In the remainder of this letter we introduce a coarse-grained model designed to reproduce generic DNA behaviour in both single- and double-stranded states, as well as the fundamental assembly transitions. 
We then demonstrate the use of the model for DNA nanostructures by simulating a full cycle for DNA tweezers.  

A number of other coarse-grained models of DNA have been suggested in recent years. Non-helical models, with two interaction sites per nucleotide, have been applied to duplex \cite{Ouldridge2009}, hairpin \cite{Sales-Pardo2005,Kenward2009} and four-arm junction formation \cite{Ouldridge2009}, as well as the gelation of oligonucleotide functionalized colloids \cite{Starr2006}. Helical models with two \cite{Drukker2001} or three \cite{Sambriski2008, Sambriski2009} sites per nucleotide have been used to study denaturation and hybridization of double-stranded DNA.  However, to study the formation of nanostructures or the operation of hybridization-driven nanodevices,  it is essential to have a physically reasonable representation of both single and double-stranded states. Earlier models either 
neglect the helicity of double-stranded DNA, or impose it through restrictions on the backbone of a single strand, which leads to an unphysical representation of single-stranded DNA. Furthermore, the thermodynamic 
properties of hybridization (particularly the widths of transitions) have not been well reproduced. 

\begin{figure}
\includegraphics[width =8cm, angle =0]{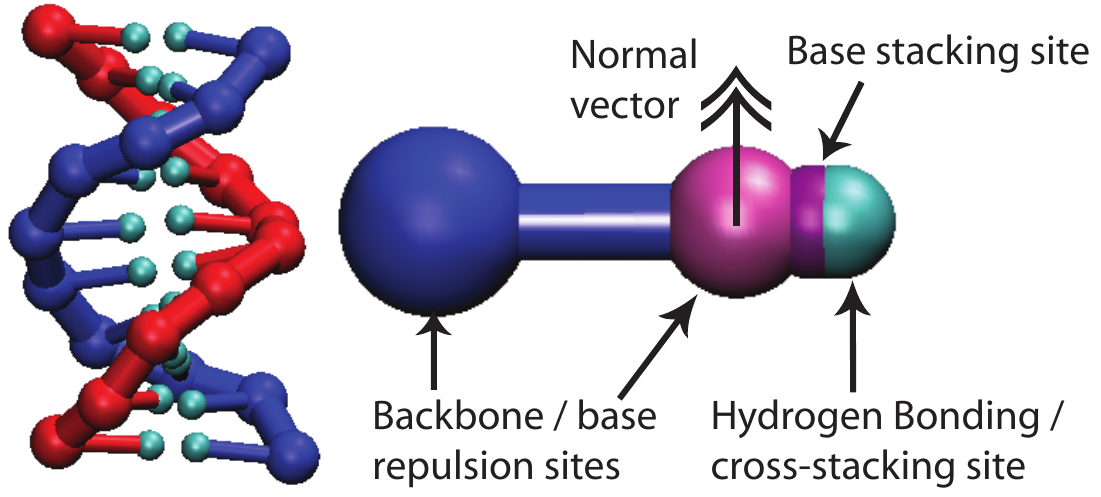}
\caption{\label{schematic} (colour online) A duplex as represented by the model, and a detailed view of a nucleotide.}
\end{figure}

We take a `top-down' approach to DNA modeling, aiming to capture the generic properties of DNA that are important for assembly rather than reproducing all of the microscopic structural details. DNA is modelled as a string of rigid nucleotides, as depicted in Fig.\ \ref{schematic}, with one interaction site to represent the backbone and three for the base. The plane of the base is indicated by an additional `normal vector'. The backbone sites are connected via FENE springs, and act as soft repulsion centres (along with the base repulsion sites) to reproduce steric interactions.

The helicity of our model results directly from the stacking interactions between base stacking interaction sites. Consecutive bases attract each other with a minimum at approximately 3.4\,\AA, shorter than the equilibrium FENE spring length of approximately 6.5\,\AA. We modulate this interaction according to the relative alignment of the normal vectors, and the alignment of the normal vectors with the inter-site vector. Thus the system is driven towards forming helical stacks of coplanar bases: right handedness is imposed by setting the attraction to zero if the bases stack left-handedly.

Hydrogen bonding is represented by an attraction between hydrogen bonding sites of complementary bases, modulated by factors favouring co-linear nucleotides with antiparallel normals. With the stacking interaction, hydrogen bonding drives the formation of right-handed double helices with the approximate geometry of B-DNA. We also include a cross-stacking interaction between bases that are diagonally opposite each other in a double helix, enabling the tendency of `dangling ends' to stabilize duplexes to be reproduced. The complete form of all potentials can be found in the Ref.\ \cite{Supplementary2009}.

For simplicity, several features of DNA have been neglected. Firstly, although only complementary bases can bond in our model, all bases are otherwise identical; at this stage we are interested in the generic properties of DNA assembly rather than specific base heterogeneity effects. Secondly, we fit the parameters  using experimental data at just one salt concentration, [Na] = 500\,mM, where the Debye screening length is short ($\sim4.5$\,\AA) and most properties are only weakly salt dependent. Finally, major and minor grooving are neglected.

We simulate the model using the `virtual move Monte Carlo' algorithm of Whitelam and Geissler \cite{Whitelam2007}. Due to the system's simplicity and the efficiency of the algorithm, denaturation and hybridization of short duplexes can be observed without biasing the ensemble. To gain accurate statistics, however, we use umbrella sampling techniques \cite{Torrie1977} to characterize the basic DNA transitions.

The simplest of these is single-stranded stacking, in which ssDNA undergoes a transition from an ordered, helical form at low temperature to a disordered structure at high temperature \cite{Saenger1984}. Our model reproduces a broad, almost uncooperative transition with an enthalpy of $\Delta H^{stack}=-5.6$\,kcal\,mol$^{-1}$ and entropy of $\Delta S^{stack}=-16.6$\,kcal\,mol$^{-1}$\,K$^{-1}$, consistent with the experiments of Holbrook {\it et al.} \cite{Holbrook1999}.

\begin{figure}
\includegraphics[width=8.5cm]{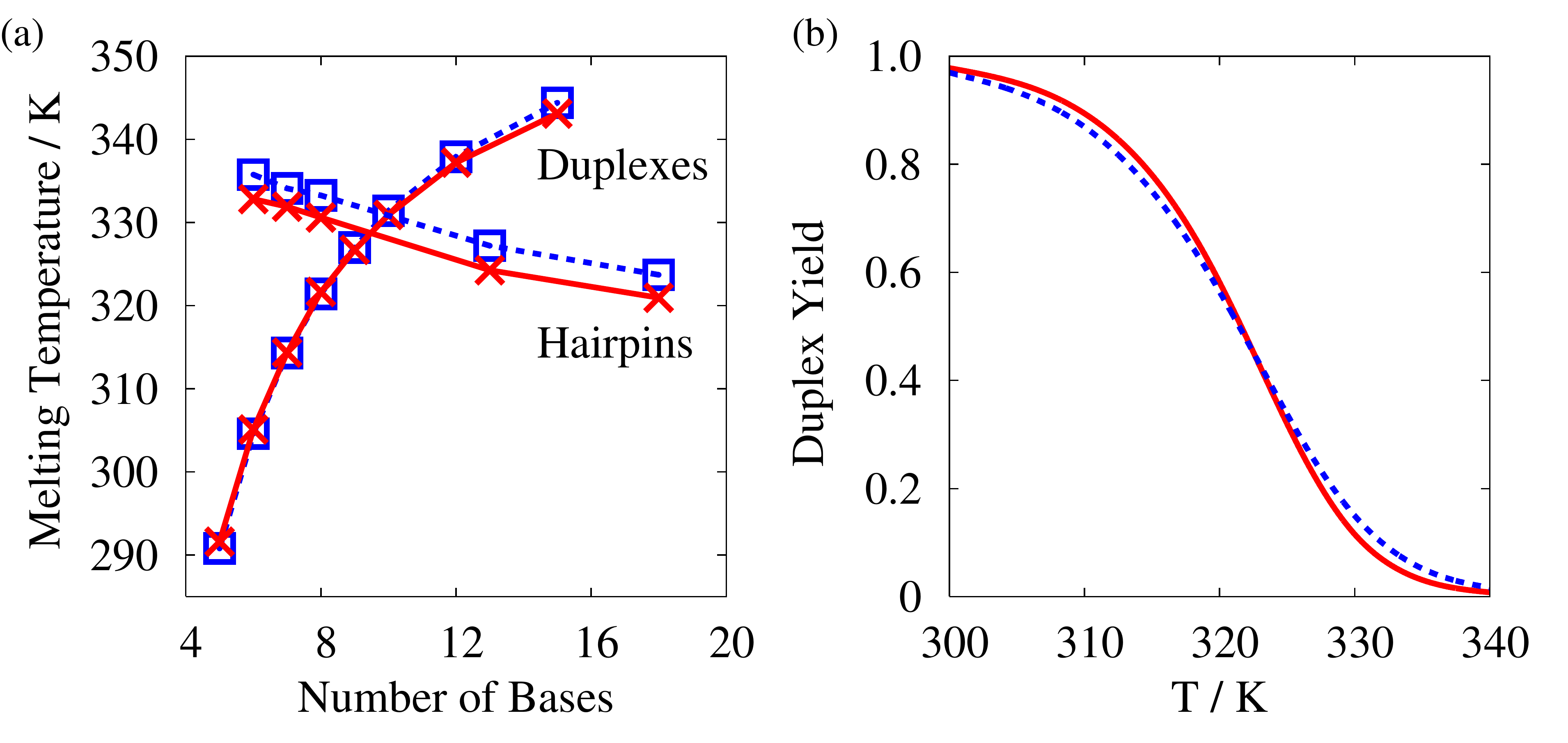}
\caption{\label{Tm}(colour online) (a) Comparison of melting temperatures as computed for our model (crosses connected by a solid line) and predicted by the nearest-neighbour model \cite{SantaLucia2004} (squares connected by a dashed line) for duplexes as a function of the single-stranded length, and hairpins as a function of loop length for a stem of six bases. (b) Melting profile for an eight base duplex as predicted by our model (solid line) and the nearest-neighbour model (dashed line).}
\end{figure}

We study duplex formation by simulating two complementary strands in a box at an effective concentration of $0.317\,{\rm mM}$, extrapolating to bulk using the method in Ref.\ \cite{Ouldridge_bulk_2009}. We compare to melting temperatures ($T_{\rm m}$) obtained from the  nearest neighbor model of SantaLucia \cite{SantaLucia2004}, which is able to accurately predict experiments, for strands consisting of `average bases' (defined by averaging over the parameters for all possible complementary base pair steps). Fig.\ \ref{Tm} shows that our model is in excellent agreement with the predictions for $T_{\rm m}$ over a range of duplex lengths. Importantly, transition widths are also consistent to within approximately 2\,K, and thus the agreement in $T_{\rm m}$ will hold over a range of concentrations.

%\begin{figure}[b]

\begin{figure*}
\includegraphics[width =18cm, angle =0]{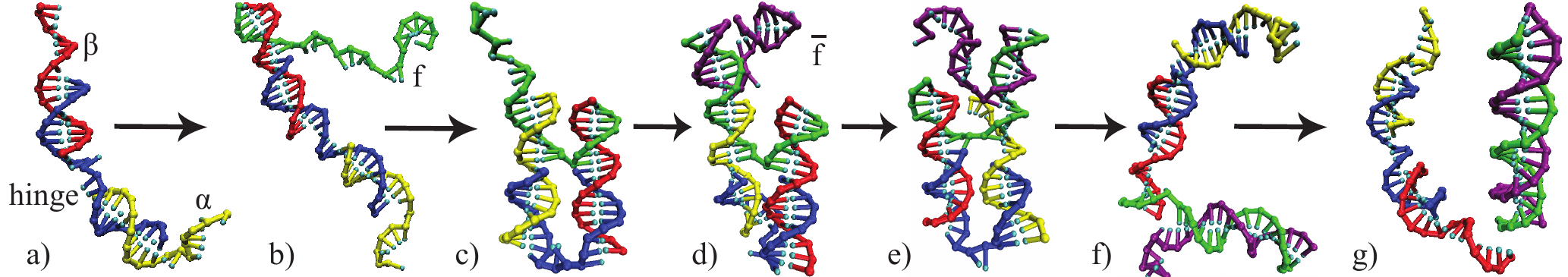}
\caption{\label{tweezers} (colour online) Simulation snapshots showing stages of operation of DNA tweezers. a) Tweezers initially open. b) Fuel (f) is added and binds to one arm ($\beta$). c) Fuel binds to the second arm ($\alpha$) and closes the tweezers. d) Antifuel (\=f) is added and binds to the toehold of the fuel. e) Antifuel begins to displace first arm of the tweezers. f) Tweezers open as first arm is displaced, and antifuel starts to displace the second arm. g) Antifuel fully hybridizes to fuel and the waste duplex is formed.}
\end{figure*}

The third basic transition is hairpin formation, in which self-complementary strands bind to themselves to form a stem and hairpin loop. Our model underestimates $T_{\rm m}$ relative to the nearest-neighbour model by approximately 3\,K (less than 1\% of the absolute temperature), but importantly captures the dependence on loop (Fig.\ \ref{Tm}) and stem length (not shown). 

In addition to thermodynamics, the model reproduces many of the physical properties of DNA essential for nanotechnology. Model duplexes have a pitch of 10.4 base pairs per turn, a persistence length of 160 base pairs and an RMSD of $3.7^{\rm o}$ in the twist of each base pair rise. Unstacked single strands are comparatively flexible, having a persistence length of 18.2\,\AA\ (we define model length scales so that the average rise per base pair at 300\,K is 3.3\,\AA).
These values compare favourably with reported experimental results of 10.5 base pairs per turn \cite{Sinden1994}, 135-150 base pairs \cite{Hagerman1988}, $3.9^{\rm o}$ \cite{Hagerman1988} and 19.4\,\AA \cite{Murphy2004}, respectively.

Having demonstrated that our model reproduces the essential physics of DNA assembly, we apply it to `DNA tweezers', a simple exampl of DNA hybridization driving conformationalchanges \cite{Yurke2000}. The cycle is shown in Fig.\ \ref{tweezers}, with the tweezer unit switching between open and closed conformations as fuel (f) and antifuel (\={f}) strands are sequentially added, producing an f\={f} duplex as waste.

For simplicity we simulate a system approximately half the size of that originally used by Yurke {\it et al.}\,\cite{Yurke2000}, with the sequences listed in Ref.\ \cite{Supplementary2009}. The tweezers themselves consist of three strands (a hinge strand and two arms ($\alpha$ and $\beta$)), forming two duplex regions of ten base pairs connected by a flexible, single-stranded hinge of four bases. At the end of the duplexes, there are overhanging single-stranded sections of eight bases. The fuel f is 24 bases in length, and is complementary to the overhanging regions of the tweezers, enabling it to bind to both and close the tweezers (Fig.\ \ref{tweezers}c). The additional eight bases provide a `toehold' for binding of the antifuel  \={f}, which is also 24 bases long and complementary to the whole of f.

The tweezers, like many DNA based machines, rely on toehold-mediated strand displacement~\cite{Bath2007}. After the addition of \={f}, the closed structure becomes metastable as the free energetic minimum of the system is an f\={f} duplex isolated from the tweezers. \={f} can bind to the toehold of f (Fig.\ \ref{tweezers}(d)): \={f} and $\alpha$ then compete for binding to the rest of f. By binding to available bases, \={f} reduces the free energy barrier for dissociation of f from $\alpha$, thereby accelerating the approach to equilibrium. Once $\alpha$ is displaced, the process is repeated with $\beta$.

\begin{figure}[b]
\includegraphics[width =8cm]{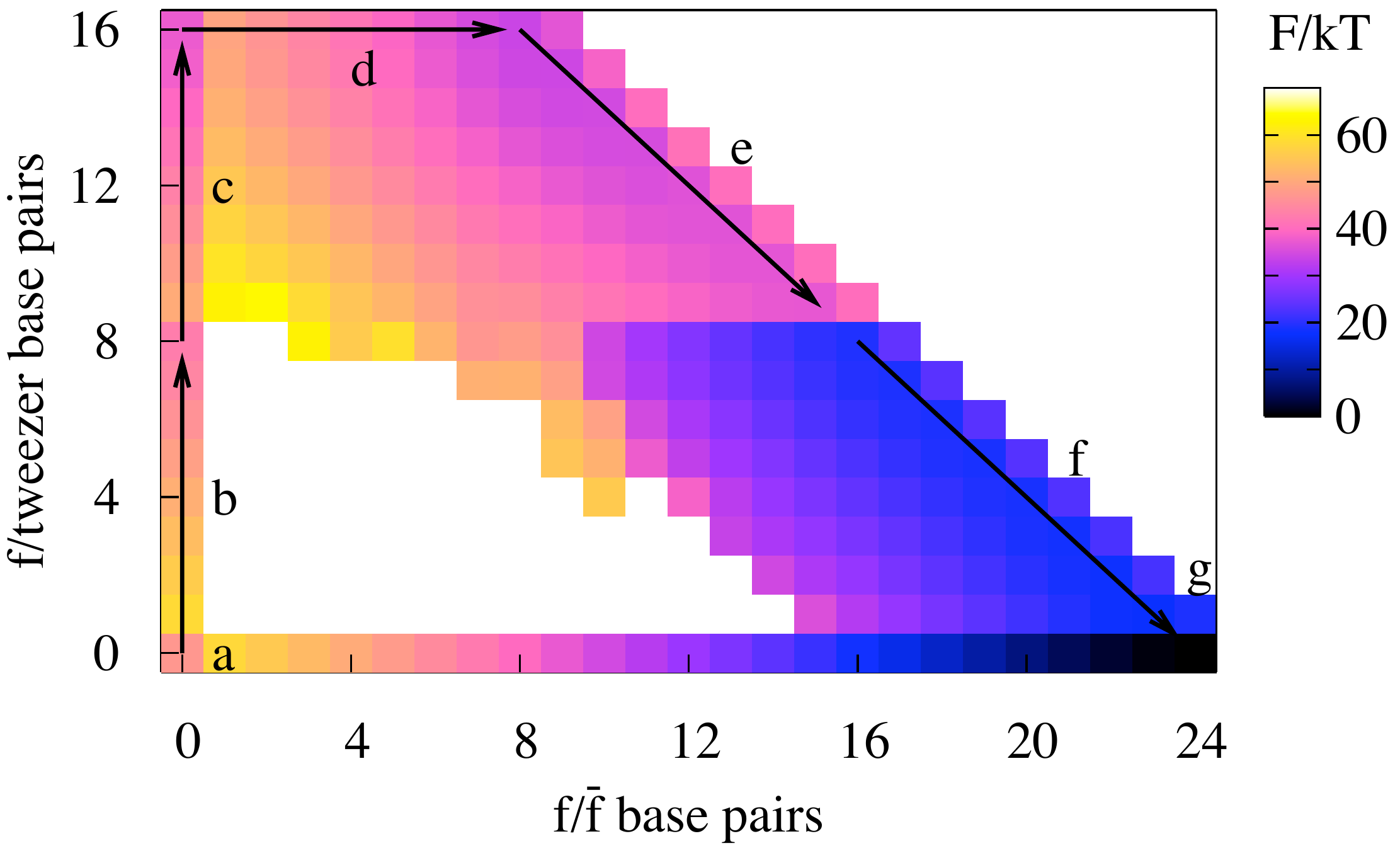}
\caption{\label{FEL} (colour online) Free energy $F$ plotted as a function the number of f/\={f}   and f/tweezer base pairs  for DNA tweezers at 300\,K. White areas indicate high free-energy regions that were unsampled.}
\end{figure}

We have sampled the free energy landscape of the system consisting of one set of tweezers and a single f and \={f}, in a periodic cell of volume  $4.19 \times 10^5$\,nm$^3$ (Fig.\ \ref{FEL}).
Every stage of the cycle is observable using unbiased simulations at 300\,K. To obtain the free energy landscape, however, we split the order parameter space into umbrella sampling windows, which were then combined using the weighted histogram analysis method \cite{Kumar1992}. Further details on how the sampling was performed are given in Ref.\ \cite{Supplementary2009}

\begin{figure}
\includegraphics[width=8.5cm]{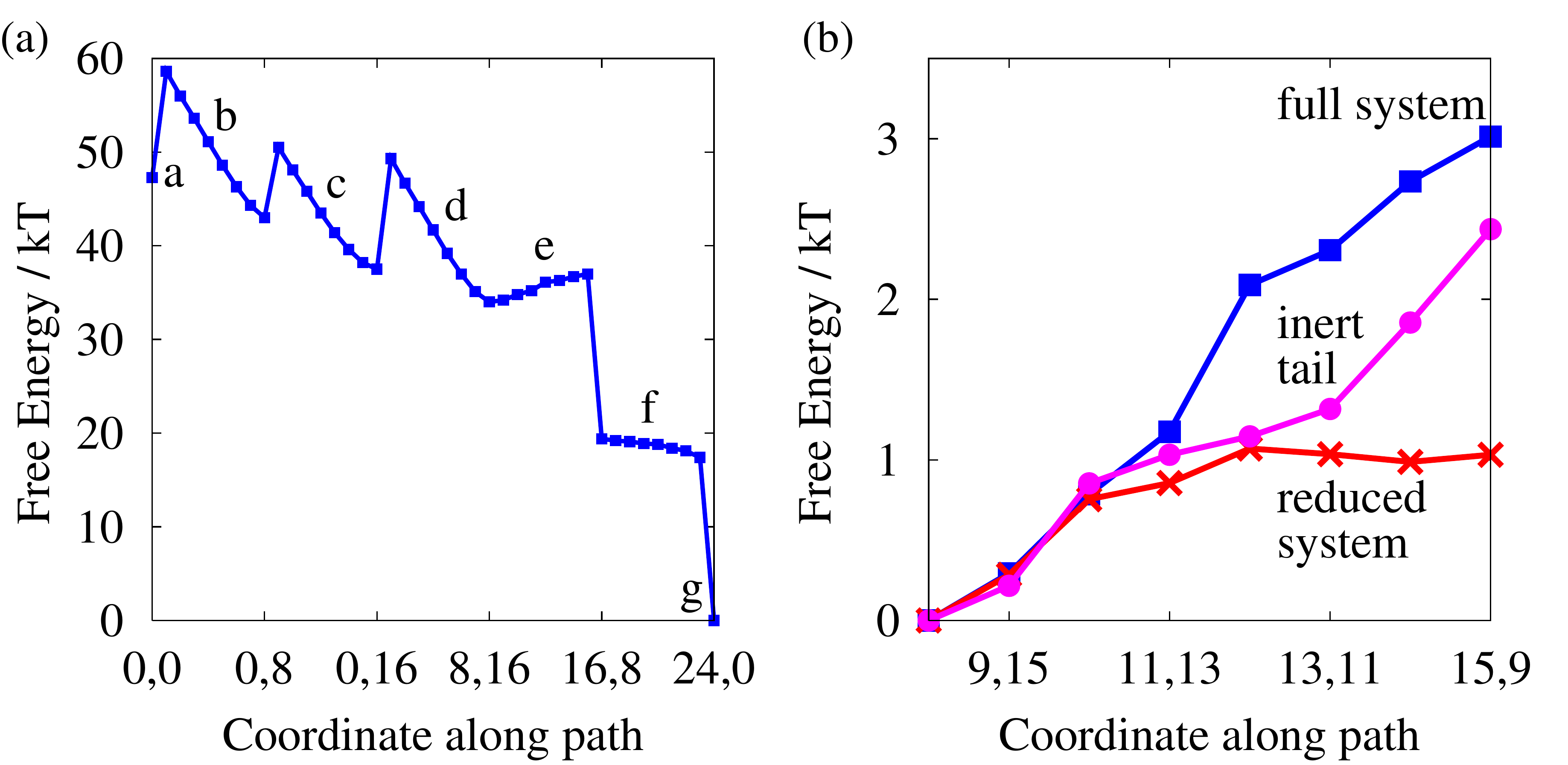}  
\caption{\label{FEP} (colour online) (a) Free energy profile along the one-dimensional pathway indicated in Fig.\ \ref{FEL}. Coordinates indicate the number of  f/\={f} and f/tweezer base pairs. (b) The displacement process `e' in more detail.  
 Squares represent the original system, circles a system with the tail of \={f} unable to form a hairpin and crosses a system with the last eight bases of \={f} and most of the f$/\beta$ arm removed (see text).}
\end{figure}

To study the cycle in detail, it is convenient to consider a one-dimensional pathway through the landscape; we use that shown by the arrows in Fig.\ \ref{FEL}. The free energy difference between `a' and `g' is  $47.12 \pm 0.21\,kT$ along this path; simulations of f and \={f} in isolation (displayed along $y= 0$ in Fig.\ \ref{FEL}) give $47.27\pm 0.11\,kT$,  the agreement supporting the accuracy of our calculations.   

The gross features of the free energy landscape are as expected. Duplex formation is highly cooperative; the pairing of two strands involves a high entropic cost for forming the first base pair, then a downhill slope in free energy as additional bonds are formed \cite{Sambriski2009}. This is reflected in Fig.\ \ref{FEP} by stages `b', `c' and `d' which essentially involve duplex formation. The large cooperativity suggests  that f will fully bind to one arm of the tweezers before binding to the second. The displacement processes (indicated by `e' and `f' in Fig.\ \ref{FEL}) are comparatively flat as the total number of interstrand base pairs is constant. Returning the tweezers to the open state (between `e' and `f') and the decoupling of the f\={f} duplex from the tweezers (`g') release the free energy stored in bringing strands together, resulting in large decreases in free energy.

Computer simulations allow for a detailed inspection of processes like displacement. Thus, Fig.\ \ref{FEP} shows that there is actually an increase in free energy of $\sim$3\,kT during the displacement of the first strand $\alpha$, even though the total number of interstrand base pairs in the system stays constant. The increase in free energy with displacement is initially steady, with a sharper jump after four bases, followed by another smooth increase.  Conversely, the displacement of the second strand $\beta$ shows a steady decrease in free energy as more bases are displaced. These slopes suggest a significant difference in speed for the two processes: our unconstrained simulations show that the first displacement requires about 10 times as many Monte Carlo moves, suggesting a slow displacement of the first arm, followed by a quicker displacement of the second.

Two effects help to explain the increase in free energy during the displacement of $\alpha$. Firstly, \={f} is capable of forming a hairpin structure, as shown in Fig.\ \ref{tweezers}(d), which is marginally stable at 300\,K. After the displacement of four bases of $\alpha$, however, the hairpin can no longer form, leading to the observed step up in free energy.  Simulations were performed in which the final eight bases of \={f} were prevented from forming hairpins (Fig.\ \ref{FEP}(b)). These show no equivalent effect, confirming this explanation. 

Unless displacing strands are deliberately designed otherwise, it is likely that small hairpins will form, with the probability of accidental hairpins increasing with the length of the strand. The nearest-neighbour model of SantaLucia \cite{SantaLucia2004} predicts that hairpins with stems of three base pairs and short loops are marginally stable at 300\,K, supporting the suggestion of our simulations that they can influence free energy profiles. Furthermore, these hairpins will form either at the start or end of displacement, when long single-stranded regions are available. As a consequence, hairpin formation will generally constitute a free energy barrier in the middle stages of displacement, thereby slowing down the process.

The second reason for the increase of free energy comes from steric effects. On binding to the toehold of f, the unbound end of \={f} has its conformational freedom restricted by the presence of the rest of the tweezers. As displacement begins, a second single-stranded region is formed, causing further steric restrictions. As more bases are displaced, the single-stranded regions are drawn into the body of the tweezers, causing additional steric restriction as illustrated in Fig.\ \ref{tweezers}(e). Computer simulations of a reduced system in which the final eight bases of \={f} (which are not involved in displacing $\alpha$) and all but the first base pair of the f/$\beta$ duplex were removed (details in Ref.\ \cite{Supplementary2009}) show a significantly flatter landscape after the initial penalty for forming two single-stranded regions, confirming this explanation (Fig.\ \ref{FEP}(b)). 
By contrast, the displacement of $\beta$ by \={f} reduces the amount of steric clashes as the tweezer unit is further separated from the f and \={f} strands with each step, leading to a decrease in free energy during the displacement.

Many of the features of the free-energy landscape --- the sharp initial rise upon forming the first base pairs, or even the more subtle effects of hairpin formation and excluded volume on the displacement steps --- are sufficiently generic that they would survive even if much more chemical detail was included in the simulations.   Future model development will include the addition of base heterogeneity effects and the explicit effects of salt concentration, but even at the current level we believe that our model will be particularly useful to study the design and operation of DNA nanomachines.  Furthermore, we anticipate many potential applications for biologically relevant rearrangement transitions, such as the 
formation of cruciform DNA \cite{Sinden1994}.

In summary, we have introduced a new coarse-grained model of DNA which reproduces its thermodynamic and structural properties, representing single-stranded stacking, duplex and hairpin transitions consistently for the first time. The model makes possible the simulation of DNA nanostructure assembly and nanomachine operation, and has the potential to be extended into the biological domain.

\end{document}